# Towards a new crown indicator: Some theoretical considerations

Ludo Waltman[1], Nees Jan van Eck, Thed N. van Leeuwen,
Martijn S. Visser, and Anthony F.J. van Raan

Centre for Science and Technology Studies, Leiden University, The Netherlands
{waltmanlr, ecknjpvan, leeuwen, visser, vanraan}@cwts.leidenuniv.nl

The crown indicator is a well-known bibliometric indicator of research performance developed by our institute. The indicator aims to normalize citation counts for differences among fields. We critically examine the theoretical basis of the normalization mechanism applied in the crown indicator. We also make a comparison with an alternative normalization mechanism. The alternative mechanism turns out to have more satisfactory properties than the mechanism applied in the crown indicator. In particular, the alternative mechanism has a so-called consistency property. The mechanism applied in the crown indicator lacks this important property. As a consequence of our findings, we are currently moving towards a new crown indicator, which relies on the alternative normalization mechanism.

## 1. Introduction

It is well known that in some scientific fields the average number of citations per publication (within a certain time period) is much higher than in other scientific fields. This is due to differences among fields in the average number of cited references per publication, the average age of cited references, and the degree to which references from other fields are cited. In addition, bibliographic databases such as Web of Science and Scopus cover some fields more extensively than others (e.g., Moed, 2005). Clearly, other things equal, one will find a higher average number of citations per publication in fields with a high database coverage than in fields with a low database coverage.

In citation-based research performance evaluations, it is crucial that one carefully controls for the above-mentioned differences among fields. This is especially the case for performance evaluations at higher levels of aggregation, such as at the level of countries, universities, or multi-disciplinary research groups. In performance evaluation studies, our institute, the Centre for Science and Technology Studies (CWTS) of Leiden University, uses a standard set of bibliometric indicators (Van Raan, 2005). Our best-known indicator, which we usually refer to as the crown indicator, relies on a normalization mechanism that aims to correct for the above-mentioned differences among fields. An indicator similar to the crown indicator is used by the Centre for R&D Monitoring (ECOOM) in Leuven, Belgium. ECOOM calls its indicator the normalized mean citation rate (e.g., Glänzel, Thijs, Schubert, & Debackere, 2009).

The normalization mechanism of the crown indicator basically works as follows. Given a set of publications, we count for each publication the number of citations it has received. We also determine for each publication its expected number of citations. The expected number of citations of a publication equals the average number of citations of all publications of the same document type (i.e., article, letter, or review) published in the same field and in the same year. To obtain the crown indicator, we divide the sum of the actual number of citations of all publications by the sum of the expected number of citations of all publications.

[1] Corresponding author.

The normalization mechanism of the crown indicator has been criticized by Lundberg (2007) and by Opthof and Leydesdorff (2010).[2] These authors have argued in favor of an alternative mechanism. According to the alternative mechanism, one first calculates for each publication the ratio of its actual number of citations and its expected number of citations and one then takes the average of the ratios that one has obtained. Lundberg refers to an indicator that uses this mechanism as the item-oriented field-normalized citation score average. This indicator is used by Karolinska Institutet in Sweden (Rehn & Kronman, 2008). Similar indicators are used by Science-Metrix in the US and Canada (e.g., Campbell, Archambault, & Côté, 2008, p. 12), by the SCImago research group in Spain (SCImago Research Group, 2009), and by Wageningen University in the Netherlands (Van Veller, Gerritsma, Van der Togt, Leon, & Van Zeist, 2009). Sandström also used a similar indicator in various bibliometric studies (e.g., Sandström, 2009, p. 33–34).

In this paper, we present a theoretical comparison between the normalization mechanism of the crown indicator and the alternative normalization mechanism discussed by Lundberg (2007) and others. We first consider two fictitious examples that provide some insight into the difference between the mechanisms. We then study the consistency (Waltman & Van Eck, 2009a, 2009b) of the mechanisms. We also pay some attention to the way in which overlapping fields should be handled. The main finding of the paper is that the alternative normalization mechanism has a more solid theoretical basis than the normalization mechanism currently applied in the crown indicator. As a consequence of this finding, CWTS is currently moving towards a new crown indicator, which relies on the alternative mechanism. For an extensive empirical comparison between the two normalization mechanisms, we refer to Waltman, Van Eck, Van Leeuwen, Visser, and Van Raan (2010).

## 2. Definitions of indicators

In this section, we provide formal mathematical definitions of the CPP/FCSm indicator and of the MNCS indicator. The CPP/FCSm indicator has been used as the crown indicator of CWTS for more than a decade. The MNCS indicator, where MNCS is an acronym for mean normalized citation score, is the new crown indicator that CWTS is planning to adopt. The two indicators differ from each other in the normalization mechanism they use. Throughout this paper, we focus on the issue of normalization for differences among fields. We do not consider the issue of normalization for differences among document types or for differences among publications of different ages. However, at the end of the paper, we will briefly comment on the latter issue. Furthermore, in the definitions provided below, there are a number of issues that we do not take into account. These issues include the document types that are considered, the way in which citation windows are chosen, and the way in which self-citations are handled. Although these issues can be of significant importance in practical research performance evaluation studies, they are not important in the theoretical setting of this paper.

[2] See also our reply to Opthof and Leydesdorff (Van Raan, Van Leeuwen, Visser, Van Eck, & Waltman, 2010) and some other contributions to the discussion (Bornmann, 2010; Moed, 2010b; Spaan, 2010).

Consider a set of $n$ publications, denoted by 1, …, $n$. Let $c_i$ denote the number of citations of publication $i$, and let $e_i$ denote the expected number of citations of publication $i$ given the field in which publication $i$ has been published. In other words, $e_i$ equals the average number of citations of all publications published in the same field as publication $i$. The field in which a publication has been published can be defined in many different ways. At CWTS, we normally define fields based on subject categories in the Web of Science database. The CPP/FCSm indicator, where CPP and FCSm are acronyms for, respectively, citations per publication and mean field citation score, is defined as

$$\text{CPP/FCSm} = \frac{\sum_{i=1}^{n} c_i \big/ n}{\sum_{i=1}^{n} e_i \big/ n} = \frac{\sum_{i=1}^{n} c_i}{\sum_{i=1}^{n} e_i}. \tag{1}$$

The CPP/FCSm indicator was introduced by De Bruin, Kint, Luwel, and Moed (1993) and Moed, De Bruin, and Van Leeuwen (1995). A similar indicator, the normalized mean citation rate, was introduced somewhat earlier by Braun and Glänzel (1990). The normalization mechanism of the CPP/FCSm indicator goes back to Schubert and Braun (1986) and Vinkler (1986).

The rationale of the CPP/FCSm indicator is as follows. Suppose the indicator is used in the evaluation of a research group. In the CPP/FCSm approach, the publications of the research group are seen as a single integrated oeuvre rather than as a number of independent works. Given the idea of a single integrated oeuvre, the distribution of citations over the individual publications in the oeuvre is not considered important. All that matters is the total number of citations received by the oeuvre. The total number of citations received by the oeuvre is in the numerator of (1). Normalization is performed with respect to the expected number of citations of the oeuvre. The expected number of citations of the oeuvre is taken to be the sum of the expected number of citations of each of the publications in the oeuvre. This is in the denominator of (1). Hence, in the CPP/FCSm approach, normalization is performed at the level of a research group's oeuvre as a whole rather than at the level of a research group's individual publications. Later on in this paper, we will point out the disadvantages of this approach. For a further discussion of the conceptual foundation of the CPP/FCSm indicator, we refer to Moed (2010b).

We now turn to the MNCS indicator. We define the MNCS indicator as[3]

$$\text{MNCS} = \frac{1}{n} \sum_{i=1}^{n} \frac{c_i}{e_i}. \tag{2}$$

The MNCS indicator is similar to the item-oriented field-normalized citation score average indicator introduced by Lundberg (2007). The normalization mechanism of the MNCS indicator is also applied in the relative paper citation rate indicator discussed by Vinkler (1996). Comparing (1) and (2), it can be seen that the CPP/FCSm indicator normalizes by calculating a ratio of averages while the MNCS indicator normalizes by calculating an average of ratios. Hence, while the CPP/FCSm indicator performs a normalization at the level of an oeuvre as a whole, the MNCS indicator performs a normalization at the level of the individual publications in an oeuvre.[4]

[3] In (1) and (2), one may in exceptional cases have a ratio with both numerator and denominator equal to zero. (Due to the way in which $c_i$ and $e_i$ are defined, it is not possible to have a ratio with a non-zero numerator and a zero denominator.) We define 0 / 0 = 1. That is, in a field with no citations, we treat all publications as having average performance.

Interestingly, the CPP/FCSm indicator can be regarded as a kind of weighted version of the MNCS indicator. To see this, notice that (1) can be rewritten as

$$\text{CPP/FCSm} = \frac{1}{n}\sum_{i=1}^{n} w_i \frac{c_i}{e_i}, \tag{3}$$

where $w_i$ is given by

$$w_i = \frac{e_i}{\sum_{j=1}^{n} e_j \Big/ n}. \tag{4}$$

Hence, like the MNCS indicator, the CPP/FCSm indicator can be written as an average of ratios. However, unlike the MNCS indicator, the CPP/FCSm indicator does not weigh all ratios equally. Instead, it gives more weight to ratios corresponding with publications that have a higher expected number of citations. In other words, fields with a high average number of citations per publication have more weight in the calculation of the CPP/FCSm indicator than fields with a low average number of citations per publication. This has also been noted by Lundberg (2007). In the calculation of the MNCS indicator, all fields have the same weight, regardless of their average number of citations per publication. We will come back to this difference between the CPP/FCSm indicator and the MNCS indicator later on in this paper.

The CPP/FCSm indicator and the MNCS indicator are both size independent. These indicators are intended to measure the average performance of a set of publications.[5] Although in performance evaluation studies one usually focuses on the average performance of a set of publications, the total performance of a set of publications can be of interest as well. A natural approach to measuring the total performance of a set of publications is to first measure the average performance of the set of publications and to then multiply the average performance by the total number of publications involved (Waltman & Van Eck, 2009a). When average performance is measured using the CPP/FCSm indicator, this approach yields

---

[4] In a somewhat different context, formulas similar to (1) and (2) were also studied by Egghe and Rousseau (1996a, 1996b). Egghe and Rousseau (1996a) refer to (1) as a globalizing quotient and to (2) as an averaging quotient. Spaan (2010) seems to contend that the use of a formula similar to (1) is always invalid, not only in bibliometrics but in all fields of science. In his view, one should always use a formula similar to (2). We believe that this position is much too extreme. Consider the following example (see also Egghe and Rousseau, 1996a). Suppose we know for each country in the European Union (EU) the gross national product (GNP) and the number of inhabitants, and suppose we want to determine the GNP per capita for the EU as a whole. Calculating the ratio of GNP and number of inhabitants for each country separately and taking the (unweighted) average of these ratios does not make much sense. Instead, one should calculate the total GNP and the total number of inhabitants of the EU and take the ratio of these two numbers. Hence, one should use a formula that is similar to (1) rather than to (2).

[5] Citation-based indicators in fact measure one specific aspect of research performance, namely the aspect of citation impact. In practice, citation-based indicators are often interpreted as approximate measures of research performance. Throughout this paper, we use the term *performance* to refer specifically to the citation impact of publications rather than to research performance in general.

$$\text{CPP/FCSm}\times n = \frac{\sum_{i=1}^{n} c_i}{\sum_{i=1}^{n} e_i / n}. \quad (5)$$

At CWTS, we refer to the indicator in (5) as the brute force indicator. This indicator is for example used in our Leiden Ranking of universities (CWTS, n.d.). When instead of the CPP/FCSm indicator the MNCS indicator is used for measuring average performance, one obtains

$$\text{TNCS} = \text{MNCS} \times n = \sum_{i=1}^{n} \frac{c_i}{e_i}. \quad (6)$$

We refer to this indicator as the TNCS indicator, where TNCS is an acronym for total normalized citation score. The TNCS indicator is similar to what Lundberg (2007) refers to as the total field-normalized citation score indicator.

## 3. Example 1

The following fictitious example provides some insight into the difference between the CPP/FCSm indicator and the MNCS indicator. Suppose we want to compare the performance of two research groups, research group A and research group B. Both research groups are active in the same field. This field consists of two subfields, subfield X and subfield Y. Research groups A and B have the same number of publications, and they both have half of their publications in subfield X and half of their publications in subfield Y. The number of publications and citations of the two research groups in the two subfields is reported in Table 1. For each subfield, the expected number of citations of a publication is also reported in the table.

Table 1. Number of publications (P) and citations (C) of research groups A and B in subfields X and Y.

| | Expected cit. per pub. | Research group A | Research group B |
|---|---|---|---|
| Subfield X | 10 | P = 100, C = 1000 | P = 100, C = 2200 |
| Subfield Y | 20 | P = 100, C = 4000 | P = 100, C = 2400 |

As can be seen in Table 1, research group B outperforms research group A in subfield X while research group A outperforms research group B in subfield Y. The question that we want to answer is which research group has a higher overall performance. The CPP/FCSm indicator and the MNCS indicator turn out to answer this question differently.

According to the CPP/FCSm indicator, the overall performance of research group A is higher than the overall performance of research group B. This is shown in Table 2. Values of the CPP/FCSm indicator for each subfield separately are also shown in the table. Notice that research group B’s performance in subfield X is higher than research group A’s performance in subfield Y and that research group B’s performance in subfield Y is higher than research group A’s performance in subfield X. Despite of this, the CPP/FCSm indicator states that research group B has a lower overall performance than research group A.

Table 2. Values of the CPP/FCSm indicator for research groups A and B.

| | Research group A | Research group B |
|---|---|---|
| Subfield X | 1.00 | 2.20 |
| Subfield Y | 2.00 | 1.20 |
| Both subfields together | 1.67 | 1.53 |

According to the MNCS indicator, the overall performance of research group B is higher than the overall performance of research group A. This is shown in Table 3. Notice that for each subfield separately the MNCS indicator yields exactly the same results as the CPP/FCSm indicator. For both subfields together, however, the indicators yield different results. In fact, they even yield opposite rankings of the two research groups.

Table 3. Values of the MNCS indicator for research groups A and B.

| | Research group A | Research group B |
|---|---|---|
| Subfield X | 1.00 | 2.20 |
| Subfield Y | 2.00 | 1.20 |
| Both subfields together | 1.50 | 1.70 |

Why does the CPP/FCSm indicator favor research group A over research group B? This is because the CPP/FCSm indicator gives more weight to subfield Y than to subfield X while the MNCS indicator weighs both subfields equally. This difference can be seen by comparing (2) with (3) and (4) (see Section 2). The CPP/FCSm indicator and the MNCS indicator agree with each other that an appropriate measure of the performance of a single publication is the ratio of the publication's actual number of citations and the publication's expected number of citations. As indicated by (2), the MNCS indicator calculates the performance of a set of publications as an unweighted average of the performance of the individual publications in the set. Since in the case of research groups A and B the number of publications in subfield X equals the number of publications in subfield Y, the MNCS indicator weighs both subfields equally. Unlike the MNCS indicator, the CPP/FCSm indicator calculates the performance of a set of publications as a weighted average of the performance of the individual publications in the set. As indicated by (3) and (4), publications with a higher expected number of citations have a higher weight. Since in subfield Y the expected number of citations of a publication is higher than in subfield X, the CPP/FCSm indicator gives more weight to subfield Y than to subfield X. In subfield Y, research group A strongly outperforms research group B, and therefore the CPP/FCSm indicator states that research group A has a higher overall performance than research group B.

Should publications be weighed differently depending on their field, like the CPP/FCSm indicator does? In general, we do not believe this to be desirable. Indicators such as the CPP/FCSm indicator and the MNCS indicator aim to correct for differences among fields. To achieve this aim, the number of citations of a publication should be normalized for differences among fields. However, after this normalization has been performed, there seems to be no reason to treat publications from different fields differently. Instead, after normalization, publications from different fields should be treated equally. This is exactly what the MNCS indicator does. By treating publications from different fields differently, the CPP/FCSm indicator introduces a bias towards fields with high a expected number of citations.

To further illustrate this point, suppose the number of publications and citations of research groups A and B in subfields X and Y is given by Table 4 rather than by

Table 1. Notice that the only thing that has changed is that the actual and expected numbers of citations in subfield Y have been divided by four. Since both for research group A and for research group B the performance in each subfield separately has not changed, it seems natural to also expect no changes in the overall performance of the research groups. In the case of the MNCS indicator, there are indeed no changes. In the case of the CPP/FCSm indicator, however, research group A's value decreases from 1.67 to 1.33 while research group B's value increases from 1.53 to 1.87 (see Table 5). This seems a counterintuitive result. Although nothing substantive has changed, the ranking of the two research groups according to the CPP/FCSm indicator has reversed.

Table 4. Number of publications (P) and citations (C) of research groups A and B in subfields X and Y. (Modified version of Table 1.)

| | Expected cit. per pub. | Research group A | Research group B |
|---|---|---|---|
| Subfield X | 10 | P = 100, C = 1000 | P = 100, C = 2200 |
| Subfield Y | 5 | P = 100, C = 1000 | P = 100, C = 600 |

Table 5. Values of the CPP/FCSm indicator for research groups A and B. (Modified version of Table 2.)

| | Research group A | Research group B |
|---|---|---|
| Subfield X | 1.00 | 2.20 |
| Subfield Y | 2.00 | 1.20 |
| Both subfields together | 1.33 | 1.87 |

## 4. Example 2

We now turn to another fictitious example that demonstrates the difference between the CPP/FCSm indicator and the MNCS indicator. The example also illustrates the possible policy relevant consequences of the difference between the indicators. Suppose the faculty of natural sciences of some university finds itself in the following situation. The faculty is doing research in two broad fields, chemistry and physics. (For simplicity, we do not break down these fields into subfields.) When differences among fields are corrected for, the chemists and the physicists working at the faculty turn out to perform equally well. This can be seen from the second and third column of Table 6.

Table 6. Number of publications (P) and citations (C) of the chemists and the physicists in the current situation and in two future scenarios.

| | Expected cit. per pub. | Current situation | Scenario 1 | Scenario 2 |
|---|---|---|---|---|
| Chemistry | 5 | P = 100, C = 500 | P = 100, C = 900 | P = 100, C = 500 |
| Physics | 10 | P = 100, C = 1000 | P = 100, C = 1000 | P = 100, C = 1600 |

To increase the performance of the faculty, a limited amount of money is available. The faculty wants to invest this money in new equipment for either the chemists or the physicists. The new equipment is expected to increase the average performance of the publications of the faculty. The expected effect is shown in the last two columns of Table 6. Scenario 1 shows what is expected to happen if the money is invested in new equipment for the chemists, and scenario 2 shows what is expected to happen if the money is invested in new equipment for the physicists.

Taking into account that in physics the expected number of citations of a publication is twice as high as in chemistry, it seems that an investment in new equipment for the chemists is preferable over an investment in new equipment for the physicists.[6] However, if an investment decision is made based on the expected effect on the overall CPP/FCSm indicator of the faculty, the available money will be invested in new equipment for the physicists. This can be seen in Table 7. Based on this, it seems that the CPP/FCSm indicator does not reflect the effects of the two investment opportunities in a completely satisfactory way.

Table 7. Values of the CPP/FCSm indicator in the current situation and in two future scenarios.

| | Current situation | Scenario 1 | Scenario 2 |
|---|---|---|---|
| Chemistry | 1.00 | 1.80 | 1.00 |
| Physics | 1.00 | 1.00 | 1.60 |
| Both fields | 1.00 | 1.27 | 1.40 |

Suppose now that an investment decision is made based on the expected effect on the overall MNCS indicator of the faculty. As can be seen in Table 8, the available money will then be invested in new equipment for the chemists. Given the information that is available, this indeed seems the best decision.

Table 8. Values of the MNCS indicator in the current situation and in two future scenarios.

| | Current situation | Scenario 1 | Scenario 2 |
|---|---|---|---|
| Chemistry | 1.00 | 1.80 | 1.00 |
| Physics | 1.00 | 1.00 | 1.60 |
| Both fields | 1.00 | 1.40 | 1.30 |

Why does the CPP/FCSm indicator favor an investment in new equipment for the physicists over an investment in new equipment for the chemists? This is again due to the bias of the CPP/FCSm indicator towards fields with a high expected number of citations. In the above example, publications in physics have a higher expected number of citations than publications in chemistry. In the calculation of the CPP/FCSm indicator, publications in physics are therefore overweighted compared with publications in chemistry. As a consequence, a relatively small increase in the performance of publications in physics can lead to a relatively large increase of the CPP/FCSm indicator.

## 5. Consistency of indicators

In this section, we study the consistency of our indicators of interest. Consistency is a mathematical property that a bibliometric indicator may or may not have. In earlier research (Waltman & Van Eck, 2009a, 2009b), it was pointed out that the well-known *h*-index (Hirsch, 2005) does not have the property of consistency.

[6] An investment in new equipment for the physicists yields a larger increase in the absolute number of citations of the faculty than an investment in new equipment for the chemists (600 vs 400). However, when looking at the relative number of citations (i.e., the number of citations after correcting for field differences), an investment in new equipment for the chemists has a larger effect than an investment in new equipment for the physicists (400 / 5 = 80 vs 600 / 10 = 60).

We first introduce some mathematical notation. For our purpose, a publication can be represented by an ordered pair $(c, e)$, where $c$ and $e$ denote, respectively, the actual and expected number of citations of the publication. $c$ is a non-negative integer, and $e$ is a non-negative rational number. Moreover, if $e$ equals zero, $c$ must equal zero as well. In other words, in a field in which the average number of citations per publication equals zero, all publications must have zero citations. We define $P$ as the set of all ordered pairs $(c, e)$. A set of $n$ publications can be represented by a multiset $S = \{(c_1, e_1), \ldots, (c_n, e_n)\}$, where $(c_1, e_1), \ldots, (c_n, e_n) \in P$.[7] We define $\Sigma$ as the set of all non-empty multisets $S$. Hence, $\Sigma$ denotes the set of all non-empty sets of publications. We define a bibliometric indicator as a function from $\Sigma$ to the set of non-negative rational numbers.

We make a distinction between on the one hand consistency of indicators of the average performance of a set of publications and on the other hand consistency of indicators of the total performance of a set of publications. We first consider the latter type of consistency. We define this type of consistency as follows.

**Definition 1.** Let $f_T$ denote a bibliometric indicator of the total performance of a set of publications. $f_T$ is said to have the property of *consistency* if

$$f_T(S_1) \geq f_T(S_2) \Leftrightarrow f_T(S_1 \cup \{(c,e)\}) \geq f_T(S_2 \cup \{(c,e)\}) \tag{7}$$

for all $S_1, S_2 \in \Sigma$ and all $(c, e) \in P$.[8]

Informally, the definition states that an indicator of total performance is consistent if adding the same publication to two different sets of publications never changes the way in which the indicator ranks the sets of publications relative to each other. This idea of consistency was also discussed by Waltman and Van Eck (2009a, 2009b). A similar idea was discussed by Marchant (2009a, 2009b), who referred to it as independence rather than consistency.

It seems very natural to expect that an indicator of total performance is consistent. It can be readily seen that the TNCS indicator defined in (6) is indeed consistent. However, the brute force indicator defined in (5) is not consistent. To see this, consider the following example. Let $S_1 = \{(3, 1)\}$ and $S_2 = \{(12, 6)\}$, and suppose a publication with $(c, e) = (0, 2)$ is added to both $S_1$ and $S_2$. Before adding the publication, the brute force indicator has a value of 3 for $S_1$ and 2 for $S_2$. After adding the publication, the brute force indicator has a value of 2 for $S_1$ and 3 for $S_2$. Hence, adding the same publication to both $S_1$ and $S_2$ causes a reversal of the way in which the brute force indicator ranks the two sets of publications. This shows the inconsistency of the brute force indicator.[9]

[7] Since $S$ is a multiset rather than an ordinary set, the elements of $S$ need not be unique. Hence, it is possible that $(c_i, e_i) = (c_j, e_j)$ for $i \neq j$. In other words, it is possible to have multiple publications with the same actual and expected number of citations.

[8] Let $S = \{(c_1, e_1), \ldots, (c_n, e_n)\} \in \Sigma$ and $T = \{(c_{n+1}, e_{n+1}), \ldots, (c_{n+m}, e_{n+m})\} \in \Sigma$. Throughout this paper, we define $S \cup T = \{(c_1, e_1), \ldots, (c_{n+m}, e_{n+m})\}$. We further note that $\alpha \geq \beta \Leftrightarrow \gamma \geq \delta$ for all rational numbers $\alpha$, $\beta$, $\gamma$, and $\delta$ implies that $\alpha > \beta \Leftrightarrow \gamma > \delta$, that $\alpha = \beta \Leftrightarrow \gamma = \delta$, and that $\alpha < \beta \Leftrightarrow \gamma < \delta$. Hence, it implies that the ranking of $\alpha$ and $\beta$ relative to each other is the same as the ranking of $\gamma$ and $\delta$ relative to each other.

We now turn to consistency of indicators of the average performance of a set of publications. For indicators of average performance, we use a slightly different definition of consistency than for indicators of total performance.

**Definition 2.** Let $f_{\mathrm{A}}$ denote a bibliometric indicator of the average performance of a set of publications. $f_{\mathrm{A}}$ is said to have the property of *consistency* if

$$f_{\mathrm{A}}(S_1) \geq f_{\mathrm{A}}(S_2) \Leftrightarrow f_{\mathrm{A}}(S_1 \cup \{(c,e)\}) \geq f_{\mathrm{A}}(S_2 \cup \{(c,e)\}) \qquad (8)$$

for all $S_1, S_2 \in \Sigma$ such that $|S_1| = |S_2|$ and all $(c, e) \in P$.[10]

According to this definition, an indicator of average performance is consistent if adding the same publication to two different but equally large sets of publications never changes the way in which the indicator ranks the sets of publications relative to each other. A similar idea, referred to as independence rather than consistency, was discussed by Bouyssou and Marchant (2010).

Like for indicators of total performance, consistency also seems an appealing property for indicators of average performance. It is not difficult to see that the MNCS indicator indeed has the property of consistency. The CPP/FCSm indicator, however, does not have this property. This can be seen using the same example as given above for the brute force indicator. In this example, adding the same publication to both $S_1$ and $S_2$ causes the value of the CPP/FCSm indicator to decrease from 3 to 1 for $S_1$ and from 2 to 3/2 for $S_2$. Hence, adding the publication leads to a reversal of the ranking of $S_1$ and $S_2$ relative to each other. This violates the property of consistency.

Let us introduce another property that an indicator of average performance may or may not have. This is an almost trivial property. We refer to it as the property of homogeneous normalization.

**Definition 3.** Let $f_{\mathrm{A}}$ denote a bibliometric indicator of the average performance of a set of publications. $f_{\mathrm{A}}$ is said to have the property of *homogeneous normalization* if

$$f_{\mathrm{A}}(S) = \frac{\sum_{i=1}^{n} c_i \big/ n}{e} \qquad (9)$$

for all $S = \{(c_1, e_1), \ldots, (c_n, e_n)\} \in \Sigma$ such that $e_1 = \ldots = e_n = e$.[11]

The property of homogeneous normalization is concerned with homogeneous sets of publications. In the context of this paper, a homogeneous set of publications is a set of publications that all belong to the same field. According to the above definition, an indicator of average performance has the property of homogeneous normalization if, in the case of a set of publications that all belong to the same field, the indicator equals the average number of citations per publication divided by the field's expected number of citations per publication. Homogeneous normalization seems to be a very natural property for an indicator of average performance. It can be readily seen that both the CPP/FCSm indicator and the MNCS indicator have the property of homogeneous normalization.

[9] Notice also that adding the publication leads to a decrease of the value of the brute force indicator for $S_1$. It seems natural to expect that an indicator of total performance never decreases when a publication is added (Waltman & Van Eck, 2009a). However, as the example shows, the brute force indicator does not have this property.

[10] We use $|S|$ to denote the number of elements of (multi)set $S$.

[11] Like in (1) and (2), we define $0 / 0 = 1$ in (9).

The MNCS indicator has both the property of homogeneous normalization and the property of consistency. An obvious question is whether there are any other indicators of average performance that also have both of these properties. The following theorem provides a negative answer to this question.

**Theorem 1.** The MNCS indicator is the only bibliometric indicator of the average performance of a set of publications that has both the property of homogeneous normalization and the property of consistency.

A proof of the theorem is provided in the appendix. It follows from the theorem that the only direct alternative to the CPP/FCSm indicator that has the property of consistency is the MNCS indicator. There are other consistent indicators,[12] but since these indicators do not have the property of homogeneous normalization, they cannot be seen as direct alternatives to the CPP/FCSm indicator.

## 6. How to handle overlapping fields?

In the previous sections, we have shown that the MNCS indicator has attractive theoretical properties. In this section, we therefore focus exclusively on the MNCS indicator. We study how the indicator should be calculated in the case of overlapping fields.

A nice property that we would like the MNCS indicator to have is that the indicator has a value of one when calculated for the set of all publications published in all fields. If there are no publications that belong to more than one field, it is easy to see that the MNCS indicator indeed has this property. However, at CWTS we normally define fields based on subject categories in the Web of Science database and these subject categories are overlapping. Many publications therefore belong to more than one field. Special care then needs to be taken to ensure that the MNCS indicator has the above-mentioned property.

Consider the following example. Suppose the scientific universe consists of just three fields, field X, field Y, and field Z, and suppose just five publications have been published in these fields during a certain time period. For each publication, the field in which it has been published as well as the number of citations it has received is listed in Table 9. Notice that publication 5 belongs both to field X and to field Y. Hence, fields X and Y are overlapping.

Table 9. Overview for each publication of the field in which it has been published and the number of citations it has received.

| | Field | Citations |
|---|---|---|
| Publication 1 | X | 2 |
| Publication 2 | X | 3 |
| Publication 3 | Y | 8 |
| Publication 4 | Z | 6 |
| Publication 5 | X and Y | 5 |

[12] A trivial example is given by $f_A(S) = 1$ for all $S \in \Sigma$.

Because publications 1, 2, 3, and 4 each belong to only one field, it is straightforward to calculate their expected number of citations. Publications 1 and 2 belong to field X, and their expected number of citations therefore equals the average number of citations of all publications published in field X. This yields

$$e_1 = e_2 = \frac{2 + 3 + 5/2}{1 + 1 + 1/2} = 3. \tag{10}$$

As can be seen, publication 5 has a weight of 1/2 in this calculation. This is because publication 5 belongs half to field X and half to field Y. The expected number of citations of publication 3 is given by

$$e_3 = \frac{8 + 5/2}{1 + 1/2} = 7, \tag{11}$$

where publication 5 again has a weight of 1/2. Obviously, for publication 4 we obtain $e_4 = 6$.

How should the expected number of citations of publication 5 be calculated? One approach is to take the arithmetic average of (10) and (11). This results in $e_5 = 5$. Calculating the value of the MNCS indicator for the set of all publications published in all fields, we then obtain

$$\text{MNCS} = \frac{1}{5}\left(\frac{2}{3} + \frac{3}{3} + \frac{8}{7} + \frac{6}{6} + \frac{5}{5}\right) = \frac{101}{105}. \tag{12}$$

Notice that the MNCS indicator does not have a value of one. This means that the property formulated at the beginning of this section is violated. Because of this, calculating the expected number of citations of publication 5 by taking the arithmetic average of (10) and (11) does not seem a completely satisfactory approach.

We now discuss an alternative approach that does yield satisfactory results. The calculations in (10) and (11) are based on the idea that publication 5 belongs half to field X and half to field Y. The same idea can also be applied in the calculation of the MNCS indicator. This results in

$$\text{MNCS} = \frac{1}{5}\left(\frac{2}{3} + \frac{3}{3} + \frac{8}{7} + \frac{6}{6} + \frac{1}{2}\frac{5}{3} + \frac{1}{2}\frac{5}{7}\right) = 1. \tag{13}$$

In this case, the MNCS indicator does have the desired value of one. An equivalent way to obtain this result is to calculate the expected number of citations of publication 5 as the harmonic (rather than the arithmetic) average of (10) and (11). We then have

$$e_5 = \frac{2}{1/3 + 1/7} = \frac{21}{5}, \tag{14}$$

which gives

$$\mathrm{MNCS} = \frac{1}{5}\left(\frac{2}{3} + \frac{3}{3} + \frac{8}{7} + \frac{6}{6} + \frac{5}{21/5}\right) = 1\,. \tag{15}$$

The use of harmonic averages ensures that the MNCS indicator always has a value of one when calculated for the set of all publications published in all fields. This therefore seems the most appropriate approach to deal with overlapping fields. The approach leads to a convenient interpretation of the MNCS indicator. When the indicator has a value above one, one's publications on average perform above world average. When the indicator has a value below one, one's publications on average perform below world average. As shown above, this interpretation is not valid when arithmetic rather than harmonic averages are used in the calculation of the MNCS indicator.[13]

## 7. Discussion and conclusion

We have presented a theoretical comparison between two normalization mechanisms for bibliometric indicators of research performance (for an empirical comparison, see Waltman et al., 2010). One normalization mechanism is implemented in the CPP/FCSm indicator, also referred to at CWTS as the crown indicator. The other normalization mechanism is implemented in what we call the MNCS indicator. The examples that we have given show that the CPP/FCSm indicator sometimes yields counterintuitive results, which is not the case for the MNCS indicator. The counterintuitive results of the CPP/FCSm indicator are due to the unequal weighing of publications from different fields. Unlike the MNCS indicator, the CPP/FCSm indicator gives more weight to publications from fields with a high expected number of citations. We have also studied the consistency of both the CPP/FCSm indicator and the MNCS indicator. Consistency is a mathematical property based on the idea that the ranking of two units relative to each other should not change when both units make the same progress in terms of publications and citations. As we have pointed out, the MNCS indicator is consistent whereas the CPP/FCSm indicator is not. This is another reason why we consider the MNCS indicator preferable over the CPP/FCSm indicator. Finally, we have discussed how overlapping fields should be dealt with in the case of the MNCS indicator. Contrary to what one might expect, harmonic rather than arithmetic averages should be used to calculate the expected number of citations of a publication that belongs to multiple fields.

Based on the findings reported in this paper, CWTS is currently moving towards the MNCS indicator as its new crown indicator. However, like any bibliometric indicator of research performance, the MNCS indicator has its limitations and is in general not suitable to be used in isolation. Instead, the MNCS indicator should be used in a complementary fashion with other indicators, and the focus should be on the overall picture that emerges from the various indicators. To conclude this paper, we briefly discuss a number of important limitations and disadvantages of the MNCS indicator (some of which are shared with the CPP/FCSm indicator):

- The MNCS indicator is defined as an arithmetic average. Since citation distributions tend to be highly skewed, one should be careful with the use of arithmetic averages. If in a set of publications there is one publication that has many more citations than the rest of the publications, the value of the MNCS indicator will be largely influenced by this single publication. This may be regarded as undesirable. To deal with this issue, the MNCS indicator needs to be complemented with additional information. One possibility is to provide confidence intervals. A wide confidence interval may be an indication of the presence of one or more publications with a very large number of citations and, consequently, with a large effect on the value of the MNCS indicator. Another possibility is to complement the MNCS indicator with an indicator based on the percentage of highly cited publications, for example, the percentage of publications belonging to the top 5% or the top 10% of their field (e.g., Tijssen, Visser, & Van Leeuwen, 2002). Such an indicator is robust to the presence of publications with a very large number of citations. A third possibility is to complement the MNCS indicator with an indicator of the performance of the journals in which publications have appeared.[14] Such an indicator is also more robust than the MNCS indicator.

[13] It can be shown that in the case of the CPP/FCSm indicator the situation is exactly the other way around. If arithmetic averages are used, the CPP/FCSm indicator has the property that it always has a value of one when calculated for the set of all publications published in all fields. If harmonic averages are used, the CPP/FCSm indicator does not have this property.

- In the case of normalization for the year in which a publication was published, the value of the MNCS indicator can be largely influenced by very recent publications.[15] A very recent publication (e.g., a publication that is less than one year old) usually has a rather low expected number of citations (quite close to zero in many cases). Hence, even if such a publication has been cited only one or two times, the ratio of its actual number of citations and its expected number of citations may already be quite high. Because of this, very recent publications can have a large effect on the value of the MNCS indicator. This may be regarded as undesirable. One way to deal with this issue is to complement the MNCS indicator with an indicator of the performance of the journals in which publications have appeared. Such an indicator is much less sensitive to very recent publications. Another possibility is to leave out the most recent publications in the calculation of the MNCS indicator. We investigate this possibility in detail in another paper (Waltman et al., 2010).
- The MNCS indicator requires a classification scheme for assigning publications to fields. Given the fuzziness of disciplinary boundaries and the multidisciplinary character of many research topics, such a scheme will always involve some arbitrariness and will never be completely satisfactory. This issue can be dealt with in various ways. Perhaps the most attractive approach is to complement the MNCS indicator with an indicator based on the idea of source normalization (Moed, 2010a). A source-normalized indicator aims to correct for differences among fields based on the characteristics of citing publications or citing journals. Using a source-normalized indicator, the need for a classification scheme can be avoided. Currently, however, source-normalized indicators have only been used for measuring the performance of journals (Moed, 2010a; Zitt, 2010; Zitt & Small, 2008). In our ongoing research, we are investigating the use of source-normalized indicators for measuring the performance of research groups and institutes (Waltman & Van Eck, 2010).

[14] We use the MNJS indicator for this purpose, where MNJS is an acronym for mean normalized journal score. The MNJS indicator is related to the JCSm/FCSm indicator used previously by CWTS (Van Raan, 2005) in the same way as the MNCS indicator is related to the CPP/FCSm indicator. Like the MNCS indicator, the MNJS indicator corrects for differences among fields.

[15] This is a difference with the CPP/FCSm indicator. The CPP/FCSm indicator gives a low weight to very recent publications, and these publications therefore tend to have only a small effect on the value of the indicator.

- One of the basic ideas underlying the MNCS indicator is that all publications should have the same weight in the calculation of the indicator, regardless of the field to which they belong. This seems a reasonable principle, and we believe that it is certainly more reasonable than the way in which publications are weighed in the CPP/FCSm indicator, where the emphasis is on publications from fields with a high average number of citations per publication. However, the equal weighing of publications in the MNCS indicator may not always be completely satisfactory. Suppose that the MNCS indicator is calculated for a set of publications from different fields, and suppose that in some of these fields the average amount of resources (e.g., research time or money) needed to produce a publication is significantly larger than in other fields. One may then argue that publications from fields in which a relatively large amount of resources are needed to produce a publication should have more weight in the calculation of the MNCS indicator than publications from fields in which producing a publication requires only a relatively small amount of resources. By weighing publications in such a way, the MNCS indicator would provide a kind of approximate measure of the efficiency with which resources are used. Information on differences among fields in the average amount of resources needed to produce a publication is not generally available. However, in special cases in which such information is available (e.g., Aksnes & Sivertsen, 2009), it can be used to refine the MNCS indicator.

The issues listed above are all of significant importance in the daily practice of bibliometric research performance assessment. We plan to address these issues and other closely related ones in our future research.

## Acknowledgment

We would like to thank Wolfgang Glänzel for his comments on earlier drafts of this paper.

## Appendix

In this appendix, we provide a proof of Theorem 1. Let $f_A$ denote a bibliometric indicator of the average performance of a set of publications. Let $f_A$ have both the property of homogeneous normalization and the property of consistency. We will prove that $f_A$ is the MNCS indicator defined in (2).

We will first show that

$$f_A(S \cup \{(c, e)\}) = \frac{|S| f_A(S) + f_A(\{(c, e)\})}{|S| + 1} \tag{16}$$

for all $S \in \Sigma$ and all $(c, e) \in P$. Let $\alpha$ and $\beta$ be non-negative integers and $\gamma$ and $\delta$ be positive integers such that $f_A(S) = \alpha / \gamma$ and $f_A(\{(c, e)\}) = \beta / \delta$. Since $f_A(S)$ and $f_A(\{(c, e)\})$ are non-negative rational numbers, $\alpha$, $\beta$, $\gamma$, and $\delta$ are guaranteed to exist. Let $T \in \Sigma$ denote a multiset of $|S|$ identical elements $(\alpha\delta, \gamma\delta) \in P$. Using the property of homogeneous normalization, it can be seen that

$$f_{\mathrm{A}}(T) = \frac{\alpha\delta}{\gamma\delta} = \frac{\alpha}{\gamma} = f_{\mathrm{A}}(S)\,, \tag{17}$$

$$f_{\mathrm{A}}(\{(\beta\gamma, \delta\gamma)\}) = \frac{\beta\gamma}{\delta\gamma} = \frac{\beta}{\delta} = f_{\mathrm{A}}(\{(c, e)\})\,. \tag{18}$$

Given (17) and (18), the property of consistency implies that

$$f_{\mathrm{A}}(S \cup \{(c, e)\}) = f_{\mathrm{A}}(T \cup \{(c, e)\}) = f_{\mathrm{A}}(T \cup \{(\beta\gamma, \delta\gamma)\})\,, \tag{19}$$

where the second equality follows from repeated application of the consistency property. Furthermore, due to the property of homogeneous normalization,

$$f_{\mathrm{A}}(T \cup \{(\beta\gamma, \delta\gamma)\}) = \frac{|T|\,\alpha\delta + \beta\gamma}{(|T|+1)\gamma\delta} = \frac{|S|\, f_{\mathrm{A}}(S) + f_{\mathrm{A}}(\{(c, e)\})}{|S|+1}\,. \tag{20}$$

Equation (16) is obtained by combining (19) and (20).

It is now straightforward to show that $f_{\mathrm{A}}$ is the MNCS indicator defined in (2). Mathematical induction based on (16) implies that

$$f_{\mathrm{A}}(S) = \frac{1}{n}\sum_{i=1}^{n} f_{\mathrm{A}}(\{(c_i, e_i)\}) \tag{21}$$

for all $S = \{(c_1, e_1), \ldots, (c_n, e_n)\} \in \Sigma$. It follows from the property of homogeneous normalization that (21) equals (2). Hence, $f_{\mathrm{A}}$ is the MNCS indicator defined in (2). This completes the proof of the theorem.